\documentclass[aps,pra,twocolumn,superscriptaddress,10pt]{revtex4-2}
\usepackage[utf8]{inputenc}
\usepackage{mathrsfs}
\usepackage{dsfont}
\usepackage{amsmath}
\usepackage{amssymb}
\usepackage{amsthm}
\usepackage{amsfonts}
\usepackage{amstext}
\usepackage{amsopn}
\usepackage{amsxtra}
\usepackage{mathrsfs}
\usepackage{dsfont}
\usepackage{color}
\usepackage{graphicx}
\usepackage{hyperref}
\usepackage{xurl}
\usepackage{braket}
\usepackage{enumerate}
\usepackage[capitalize]{cleveref}

\newtheorem{theorem}{Theorem}

\newcommand{\R}{\mathbb{R}}

\newcommand{\bW}{\hat{W}}%{\mathbb{W}}
\newcommand{\bV}{\hat{V}}%{\mathbb{V}}
\newcommand{\bT}{\hat{T}}%{\mathbb{T}}

\newcommand{\bL}{\hat{L}}
\newcommand{\bK}{\hat{K}}
\newcommand{\bH}{\hat{H}}%{\mathbb{H}}

\newcommand{\ii}{\infty}
\newcommand\1{{\ensuremath {\mathds 1} }}
\renewcommand\phi{\varphi}

\newcommand{\cT}{\mathcal{T}}

\newcommand{\cV}{\hat{V}}

\newcommand{\cU}{\hat{U}}

\newcommand{\cM}{\mathcal{M}}

\newcommand{\cN}{\mathcal{N}}

\newcommand{\cW}{\hat{W}}

\newcommand{\cL}{\mathcal{L}}
\newcommand\pscal[1]{{\ensuremath{\left\langle #1 \right\rangle}}}

\renewcommand{\leq}{\leqslant}
\renewcommand{\hat}{\widehat}

\newcommand{\rd}{\mathrm{d}}
\newcommand{\Corr}{\hat{F}}

\newcommand{\br}{\mathbf{r}}
\newcommand{\bj}{\mathbf{j}}

\begin{document}

\title{Geometric Time-Dependent Density Functional Theory}

\author{\'Eric~Canc\`es}
\email{eric.cances@enpc.fr}
\affiliation{CERMICS, \'Ecole des Ponts -- Institut Polytechnique de Paris and Inria, 6-8 avenue Blaise Pascal, Cité Descartes, 77455 Marne-la-Vallée, France}

\author{Th\'eo~Duez}
\email{theo.duez@enpc.fr}
\affiliation{CERMICS, \'Ecole des Ponts -- Institut Polytechnique de Paris and Fédération CNRS Bézout, 6-8 avenue Blaise Pascal, Cité Descartes, 77455 Marne-la-Vallée, France}

\author{Jari~van~Gog}
\email{jari.van\_gog@etu.sorbonne-universite.fr}
\affiliation{Laboratoire de Chimie Théorique, Sorbonne Université and CNRS, 4 place Jussieu, 75005 Paris, France}

\author{Asbj{\o}rn~B{\ae}kgaard~Lauritsen}
\email{lauritsen@ceremade.dauphine.fr}
\affiliation{CEREMADE, CNRS, Université Paris-Dauphine, PSL Research University, Place de Lattre de Tassigny, 75016 Paris, France}

\author{Mathieu~Lewin}
\email{Mathieu.Lewin@math.cnrs.fr}
\affiliation{CEREMADE, CNRS, Université Paris-Dauphine, PSL Research University, Place de Lattre de Tassigny, 75016 Paris, France}

\author{Julien~Toulouse}
\email{toulouse@lct.jussieu.fr}
\affiliation{Laboratoire de Chimie Théorique, Sorbonne Université and CNRS, 4 place Jussieu, 75005 Paris, France}

\begin{abstract}
We provide a new formulation of Time-Dependent Density Functional Theory (TDDFT) based on the geometric structure of the set of states constrained to have a fixed density. Orbital-free TDDFT is formulated using a hydrodynamics equation involving a new density-to-current functional map. In the corresponding Kohn--Sham equation, the density is reproduced using a non-local operator. Finally, we present numerical simulations for one-dimensional soft-Coulomb systems.
\end{abstract}

\maketitle

Density Functional Theory (DFT) is very successful in describing electrons at equilibrium (close to their ground states). It is the reference computational method for large molecules and solids, both in academic research and industrial applications. By contrast, Time-Dependent Density Functional Theory (TDDFT)~\cite{RunGro-84,Ullrichs-11,MarMaiNogGroRub-12} is less developed and still suffers from important limitations~\cite{Fuks-16,LacMai-NPJCM-23,DFT-22}. It is however the only viable computational approach for complex systems, and the demand for novel concepts and approximations in TDDFT is considerable. Practical applications of TDDFT include charge transfer in complex molecules~\cite{FolMauHamJayWahRagJonDimGaaSchLop-JPCA-23}, electronic response of large systems~\cite{JakLuBriLinSumGanBer-JCTC-25}, and ultrafast phenomena in attosecond physics~\cite{SatHubGioRub-NPJCM-25}. In this Letter we open a new avenue in TDDFT. Based on geometric concepts, we introduce a new exact scheme that seems to better describe systems driven out of equilibrium. We also present 1D numerical simulations that illustrate the potential power of our new approach, when used as a correction to adiabatic density functionals.

\smallskip
\paragraph*{Geometric orbital-free theory.}
The usual $N$-electron Hamiltonian is
\begin{equation}
\bH(t):=\bT+\bV_{\rm ee}+\bV_{\rm ext}(t),
\label{eq:H(t)}
\end{equation}
with $\bT :=-\sum_{j=1}^N\nabla^2_{\br_j}/2$ the kinetic-energy operator, 
$\bV_{\rm ee}$ the electron-electron interaction and $\bV_{\rm ext}(t):=\sum_{j=1}^N V_{\rm ext}(t,\br_j)$ a time-dependent external potential. The reference $N$-particle Schrödinger equation is
\begin{equation}
i\partial_t\Psi^{\rm S}(t)=\bH(t)\,\Psi^{\rm S}(t),\quad\Psi^{\rm S}(0)=\Psi_0.
\label{eq:Schrodinger}
\end{equation}
In this first section, we focus on orbital-free TDDFT, where the high-dimensional linear equation~\eqref{eq:Schrodinger} is replaced by an equivalent nonlinear equation involving only the one-particle density
\begin{equation}
\rho_\Psi(\br):=\bigg\langle\Psi,\sum_{j=1}^N\delta(\br-\br_j)\Psi\bigg\rangle.
\label{eq:density}
\end{equation}
To this end, we consider modified Schrödinger equations
\begin{equation}
i\partial_t\Psi(t)=\Big(\bH(t)+\Corr[\Psi_0,V_{\rm ext},\rho](t)\Big)\Psi(t),\quad\Psi(0)=\Psi_0,
\label{eq:Schrodinger_modified}
\end{equation}
where the correction term $\Corr[\Psi_0,V_{\rm ext},\rho](t)$ is such that $\rho_{\Psi(t)}(\br)=\rho(t,\br)$, a density prescribed in advance. The correction term is a functional of this $\rho$, of the time-dependent external potential $V_{\rm ext}$ and of the initial state $\Psi_0$ (assumed to have density $\rho(0,\br)$). To ensure causality, $\Corr[\Psi_0,V_{\rm ext},\rho](t)$ is assumed to only depend on $\rho(s,\br)$ and $V_{\rm ext}(s,\br)$ for past times $0\leq s\leq t$.

There exist many choices for $\Corr$. Based on geometric considerations, we will construct a functional for which the density $\rho^{\rm S}:=\rho_{\Psi^{\rm S}}$ of the Schrödinger $\Psi^{\rm S}$ in~\eqref{eq:Schrodinger} is the only solution to the implicit equation $\Corr[\Psi_0,V_{\rm ext},\rho]=0$. This is how~\eqref{eq:Schrodinger} can be reformulated in terms of the density only. In the next section, we consider the corresponding Kohn--Sham (KS) theory, which provides $\rho^{\rm S}$ in terms of orbitals obtained from a non-interacting Hamiltonian with a correction $\Corr$ that is non-zero for $\rho=\rho^{\rm S}$.

Standard TDDFT can be reinterpreted as the choice $\Corr^{\rm VP}[\Psi_0,V_{\rm ext},\rho](t):=\sum_{j=1}^NV(t,\br_j)$ with a potential functional $V=V[\Psi_0,V_{\rm ext},\rho]$. This comes from requiring the action to be stationary with respect to wavefunction variations at fixed density~\cite[Sec.~III.A]{MAQUI_finite-26_ppt}, which we call the Variational Principle (VP). The Runge--Gross theorem~\cite{RunGro-84} states that $V$ is unique modulo a time-dependent constant $C(t)$, hence $\rho^{\rm S}$ is the unique solution to $V[\Psi_0,V_{\rm ext},\rho]=0$. Since $V[\Psi_0,V_{\rm ext},\rho]=V[\Psi_0,0,\rho] - V_{\rm ext}$, this last condition can be formulated with the universal functional $V[\Psi_0,\rho]:=V[\Psi_0,0,\rho]$ as
\begin{equation}
V[\Psi_0,\rho](t,\br)=V_{\rm ext}(t,\br)+C(t)\quad\Longleftrightarrow \quad\rho=\rho^{\rm S}.
    \label{eq:charac_rho_S_V}
\end{equation}

\begin{figure}[t]
\includegraphics[width=0.8\columnwidth]{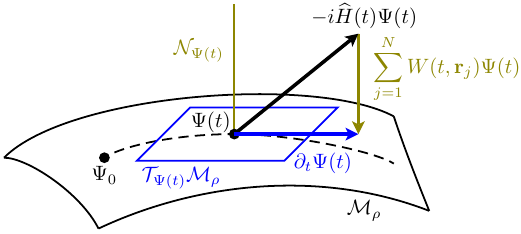}
\caption{In the Geometric Principle, the velocity $\partial_t\Psi(t)$ is the projection of $-i\hat{H}(t)\Psi(t)$ onto the tangent space $\cT_{\Psi(t)}\cM_\rho$. The normal space is given by~\eqref{eq:normal_space_Psi} and one arrives at Eq.~\eqref{eq:Schrodinger_geometric}.  \label{fig:McLachlan}}
\end{figure}

In our new Geometric Principle (GP) we do not demand the action to be stationary, but instead require that the norm of the correction, $\|\Corr^{\rm GP}[\Psi_0,V_{\rm ext},\rho]\Psi(t)\|=\|i\partial_t\Psi(t)-\hat{H}(t)\Psi(t)\|$, be the smallest it can be at all times. We first consider a prescribed \emph{time-independent} density $\rho(\br)$. The wavefunction $\Psi(t)$ in~\eqref{eq:Schrodinger_modified} then evolves on the set $\cM_\rho$ of wavefunctions having this prescribed $\rho$ and the velocity $\partial_t\Psi(t)$ must belong to the tangent space $\cT_{\Psi(t)}\cM_\rho$ at all times. As shown in Fig.~\ref{fig:McLachlan}, the correction term has minimal norm when $\partial_t\Psi(t)$ is the projection of $-i \bH(t)\Psi(t)$ onto the tangent space, thus when $-i\Corr^{\rm GP}[\Psi_0,V_{\rm ext},\rho]\Psi(t)$ is in the normal space. By simple calculations \cite[Sec.~II.B]{MAQUI_finite-26_ppt}, the normal space is
\begin{equation}
\cN_\Psi=\bigg\{\sum_{j=1}^Nw(\br_j)\Psi,\ w:\R^3\to\R\bigg\}
\label{eq:normal_space_Psi}
\end{equation}
and the new correction term is thus
\begin{equation}
\Corr^{\rm GP}[\Psi_0,V_{\rm ext},\rho]\Psi(t):=i\sum_{j=1}^NW(t,\br_j) \Psi(t),
\label{eq:form_F_non_hermitian}
\end{equation}
with a real-valued function $W(t,\br)$. For a prescribed \emph{time-dependent} density, the displacement of the tangent space in the normal direction only depends on $\partial_t\rho(t)$ and does not change the form of the correction in~\eqref{eq:form_F_non_hermitian}~\cite[Sec.~II.C]{MAQUI_finite-26_ppt}. 
By construction, $W(t,\br)$ only depends on $\Psi(t)$, $V_{\rm ext}(t)$ and $\partial_t\rho(t)$ and is thus causal. The modified Schrödinger equation \eqref{eq:Schrodinger_modified} is
\begin{equation}
i\partial_t\Psi(t)=\big(\bH(t)+i\bW(t)\big)\Psi(t),
\label{eq:Schrodinger_geometric}
\end{equation}
where $\bW(t):=\sum_{j=1}^N W(t,\br_j)$ and $W=W[\Psi_0,V_{\rm ext},\rho]$.

Equation~\eqref{eq:Schrodinger_geometric} looks like a Schrödinger equation with a complex potential, similar to those considered for describing resonances~\cite{SanCed-02,Ern-JCP-06,LiUll-08,GoyErn-11,ZhoErn-JCP-12} or open systems~\cite{GoyErnZhu-07,GebCar-PRL-04,YueRodAsp-PCCP-09}. However, here, $W$ is chosen so that $\rho_{\Psi(t)}=\rho(t)$ and thus the resulting $\Psi(t)$ stays normalized for all times. This means that $\langle \Psi(t), \hat W(t) \Psi(t) \rangle = 0$, which implies that Eq.~\eqref{eq:Schrodinger_geometric} can be rewritten in commutator form
\begin{equation}
i\partial_t\Psi(t)=\Big(\bH(t)+i\big[\bW(t)\,,\,|\Psi(t)\rangle\langle \Psi(t)|\big]\Big)\Psi(t).
\label{eq:correction_geometric}
\end{equation}
For given $\bW(t)$ and $\Psi(t)$, the modified Hamiltonian in \eqref{eq:correction_geometric} is Hermitian, and thus manifestly norm preserving.

The idea of constraining a Schrödinger dynamics using a projection on the tangent space, often attributed to McLachlan~\cite{McLachlan-64}, can be traced back to Dirac~\cite{Dirac-30} and Frenkel~\cite{Frenkel-34}. Depending on the considered constraints, the GP can lead to the same solution as the VP, or to completely different solutions~\cite{KucMeyCed-88,BroLatKesLeu-88,Raab-00,HacGuaShiHaeDemCir-20,MarBur-20,LasSu-22}. For a fixed density the GP and VP corrections and wavefunctions are very different~\cite[Secs.~IV.A, VIII.C-D]{MAQUI_finite-26_ppt}.

In~\cite[Thm.~3]{MAQUI_finite-26_ppt} we give a rigorous proof of the existence and uniqueness of $W[\Psi_0,V_{\rm ext},\rho]$ in finite dimension. In the appendix we rigorously prove a Runge--Gross theorem for Eq.~\eqref{eq:Schrodinger_geometric} with smooth potentials. Hence $W[\Psi_0,V_{\rm ext},\rho]$ is uniquely defined, when it exists, and $\rho^{\rm S}$ is the unique solution to $\Corr^{\rm GP}[\Psi_0,V_{\rm ext},\rho]=0$ or
\begin{equation}
W[\Psi_0,V_{\rm ext},\rho]=0\quad\Longleftrightarrow\quad \rho=\rho^{\rm S}.
\label{eq:orbital-free_geometric_W}
\end{equation}

The function $W$ is very different in nature from the potential $V$ used in standard TDDFT. The latter often contains high peaks and steps to force the particles to move to the desired region of space~\cite{EllFukRubMai-12}. This works due to the intricate interplay between $V$ and the kinetic operator $\hat{T}$. If we remove $\hat{T}$, then adding a real-valued potential $V$ modifies the phase of the wavefunction but not its density. In contrast, the function $W$ appears directly in the continuity equation associated with~\eqref{eq:Schrodinger_geometric} or~\eqref{eq:correction_geometric},
\begin{equation}
\partial_t\rho_{\Psi(t)}(\br)+\nabla\cdot\bj_{\Psi(t)}(\br)=2\big(\cL_{\Psi(t)}W(t)\big)(\br)
\label{eq:continuity_geometric}
\end{equation}
with the linear operator
\begin{equation}
\big(\cL_\Psi w\big)(\br)=\rho_{\Psi}(\br)w(\br)
+\int_{\R^3}\rho^{(2)}_{\Psi}(\br,\br')w(\br')\,\rd\br',
\label{eq:operator_L}
\end{equation}
where $\bj_\Psi$ and $\rho_\Psi^{(2)}$ are the current and the two-particle density of $\Psi$. The operator $\cL_\Psi$ is expected to be invertible in all cases of physical interest \cite[Sec.~VII.B]{MAQUI_finite-26_ppt}. The interpretation of~\eqref{eq:continuity_geometric} is that $W$ gives rise to sources and sinks~\cite{Ern-JCP-06}, which is a more direct and flexible way of changing the density. In fact, on a lattice, densities moving too fast are not $V$-representable due to the finite speed of propagation~\cite{LiUll-08} but can perfectly be $W$-representable \cite[Secs.~VII.C and VII.D]{MAQUI_finite-26_ppt}. 

Let us now re-interpret Eq.~\eqref{eq:orbital-free_geometric_W} in light of the continuity equation~\eqref{eq:continuity_geometric}. From the functional $W[\Psi_0,V_{\rm ext},\rho]$ we obtain a unique wavefunction $\Psi^{\rm GP}[\Psi_0,V_{\rm ext},\rho](t)$ by solving~\eqref{eq:Schrodinger_geometric} or~\eqref{eq:correction_geometric}, as well as the corresponding \emph{geometric density-to-current functional}
\begin{equation}
\bj^{\rm GP}[\Psi_0,V_{\rm ext},\rho](t,\br):=\bj_{\Psi^{\rm GP}[\Psi_0,V_{\rm ext},\rho](t)}(\br),
\label{eq:density-to-current}
\end{equation}
the current of $\Psi^{\rm GP}[\Psi_0,V_{\rm ext},\rho](t)$. It plays a central role in our new theory because, from the invertibility of $\cL_{\Psi}$ in~\eqref{eq:continuity_geometric}, we can rewrite Eq.~\eqref{eq:orbital-free_geometric_W} as 
\begin{equation}
\partial_t\rho(t,\br)+\nabla\cdot\bj^{\rm GP}\!\big[\Psi_0,V_{\rm ext},\rho\big]\!(t,\br)=0\ \Longleftrightarrow\ \rho=\rho^{\rm S}.
\label{eq:orbital-free_geometric}
\end{equation}
The sought-after $\rho^{\rm S}$ is thus the unique density satisfying the standard continuity equation with the geometric current. For $\rho=\rho^{\rm S}$, we recover the Schr\"odinger wavefunction and current. Equation~\eqref{eq:orbital-free_geometric} is our new geometric formulation of orbital-free TDDFT. It is analogous to the van Leeuwen equation~\cite{vanLeeuwen-99} in standard TDDFT. 

\smallskip
\paragraph*{Geometric Kohn--Sham theory.}
In geometric KS theory, $\rho^{\rm S}$ is reproduced using $N$ non-interacting electrons described with a Slater determinant $\Phi(t) :=(N!)^{-1/2}\det(\phi_n(t,\br_k))$, where the orbitals $\phi_n$ solve
\begin{equation}
i\partial_t\phi_n(t,\br)\!=\!\Big(-\frac{\nabla^2}2+V_{\rm ext}(t,\br)
+i\big[W^{\rm KS}(t),\gamma_{\Phi(t)}\big]\Big)\phi_n(t,\br).
\label{eq:Kohn--Sham_geometric_commutator}
\end{equation}
Here $\gamma_{\Phi(t)}(\br,\br'):=\sum_{n=1}^N \phi_n(t,\br)\phi_n(t,\br')^\dagger$ is the one-particle density matrix. The correction term is of commutator form as in~\eqref{eq:correction_geometric} with $W^{\rm KS}$ a functional to be specified so that $\rho_{\Phi(t)}= \rho^{\rm S}(t)$. To obtain a predictive scheme, this functional should not depend on the unknown $\rho^{\rm S}$.  

For an initial Slater determinant $\Phi_0$, we can apply the theory developed before with $\hat{V}_{\rm ee}=0$ and obtain the corresponding non-interacting functional $W_0[\Phi_0,V_{\rm ext},\rho]$.
If we pick $\rho=\rho^{\rm S}$ and solve~\eqref{eq:Kohn--Sham_geometric_commutator} with $W^{\rm KS}=W_0[\Phi_0,V_{\rm ext},\rho^{\rm S}]$, where $\rho_{\Phi_0}=\rho_{\Psi_0}$, we obtain the trajectory of Slater determinants reproducing $\rho^{\rm S}$. In practice we do not know $\rho^{\rm S}$ and we can only conclude that the sought-after functional $W^{\rm KS}$ in Eq.~\eqref{eq:Kohn--Sham_geometric_commutator} must coincide with $W_0[\Phi_0,V_{\rm ext},\rho^{\rm S}]$ for (and only for) $\rho_\Phi=\rho^{\rm S}$. 

Recall that in standard KS TDDFT, the universal Hartree-exchange-correlation (Hxc) potential functional~\cite[Sec.~1.3]{RugPenLee-15} reads
\begin{equation}
V_{\rm Hxc}[\Psi_0,\Phi_0,\rho]:=V_0[\Phi_0,\rho]-V[\Psi_0,\rho],
    \label{eq:V_HXC}
\end{equation}
where $V_0[\Phi_0,\rho]$ is the universal KS potential and $V[\Psi_0,\rho]$ the universal interacting potential functional. In the same spirit, we define
\begin{equation}
\widetilde W^{\rm KS}[\Psi_0,\Phi_0,V_{\rm ext},\rho]:=W_0[\Phi_0,V_{\rm ext},\rho]-W[\Psi_0,V_{\rm ext},\rho].
\label{eq:W_GKS}
\end{equation}
Let $\Phi(t)$ be the solution to the nonlinear equation~\eqref{eq:Kohn--Sham_geometric_commutator} with $W^{\rm KS}=\widetilde W^{\rm KS}[\Psi_0,\Phi_0,V_{\rm ext},\rho_\Phi]$. By uniqueness in Thm.~\ref{thm:RGW}\;(b) below, we know that this $W^{\rm KS}$ must be the one reproducing $\rho_{\Phi}$, which is by definition $W_0[\Phi_0,V_{\rm ext},\rho_{\Phi}]$. But then from~\eqref{eq:W_GKS} we obtain $W[\Psi_0,V_{\rm ext},\rho_{\Phi}]=0$ and therefore $\rho_{\Phi}=\rho^{\rm S}$ as we wanted.

The functional of Eq.~\eqref{eq:W_GKS} is a natural extension of Eq.~\eqref{eq:V_HXC} but it implicitly depends on $V_{\rm ext}$. To turn it into a universal functional we can replace $V_{\rm ext}$ by $V[\Psi_0,\rho]$ since for $\rho^{\rm S}$ we will have equality by~\eqref{eq:charac_rho_S_V}. Noticing that $W[\Psi_0,V[\Psi_0,\rho],\rho]=0$,  Eq.~\eqref{eq:W_GKS} leads to the \emph{universal geometric KS functional}
\begin{equation}
W^{\rm KS}[\Psi_0,\Phi_0,\rho]:=W_0\big[\Phi_0,V[\Psi_0,\rho],\rho\big],
\label{eq:W_universal}
\end{equation}
that is now independent of $V_{\rm ext}$. The solution $\Phi(t)$ to the nonlinear equation~\eqref{eq:Kohn--Sham_geometric_commutator} with $W^{\rm KS}=W^{\rm KS}[\Psi_0,\Phi_0,\rho_\Phi]$ has the desired density $\rho^{\rm S}$. Indeed, by uniqueness in Thm.~\ref{thm:RGW}\;(b) we know that this $W^{\rm KS}$ must be the one reproducing $\rho_\Phi$ in the external potential $V_{\rm ext}$, that is, $W_0\big[\Phi_0,V[\Psi_0,\rho_\Phi],\rho_\Phi\big]=W_0[\Phi_0,V_{\rm ext},\rho_\Phi]$. This equality means that the two potentials $V[\Psi_0,\rho_\Phi]$ and $V_{\rm ext}$ yield the same $\rho_\Phi$ in presence of this $W^{\rm KS}$. By uniqueness  of $V$ at given $W$ as stated in Thm.~\ref{thm:RGW}\;(a) below, we conclude that $V[\Psi_0,\rho_\Phi]=V_{\rm ext}+C(t)$. By~\eqref{eq:charac_rho_S_V} we finally obtain $\rho_\Phi=\rho^{\rm S}$ as desired. 

We have introduced two nonlinear functionals to be used in the geometric KS scheme~\eqref{eq:Kohn--Sham_geometric_commutator}, that reproduce the exact Schrödinger density $\rho^{\rm S}$. There are other possibilities. The choice of a given functional should be motivated by how easily approximations can be made. This is not the focus of this Letter.

For systems close to equilibrium, standard adiabatic TDDFT approximations work well. This is not the case far from equilibrium. It is thus natural to use our new geometric theory to correct the limitations of the existing adiabatic approximations. For any given approximation $V^{\rm app}_{\rm Hxc}[\Psi_0,\Phi_0,\rho]$ to the exact Hxc potential in Eq.~\eqref{eq:V_HXC}, we define the corresponding \emph{KS scheme with geometric correction} as
\begin{multline}
i\partial_t\phi_n(t,\br)=\Big(-\frac{\nabla^2}2+V_{\rm ext}(t,\br)+V^{\rm app}_{\rm Hxc}[\Psi_0,\Phi_0,\rho_\Phi](t,\br)\\
+i\big[W^{\rm corr}(t)\,,\,\gamma_{\Phi(t)}\big]\Big)\phi_n(t,\br).
\label{eq:Kohn--Sham_geometric_corrected}
\end{multline}
We can define functionals for $W^{\rm corr}$ similarly as in~\eqref{eq:W_GKS} or~\eqref{eq:W_universal}, by adding $V^{\rm app}_{\rm Hxc}[\Psi_0,\Phi_0,\rho]$ to $V_{\rm ext}$ or $V[\Psi_0,\rho]$ inside $W_0$.
If $V^{\rm app}_{\rm Hxc}$ is a good adiabatic approximation, we expect $W^{\rm corr}$ to be  small when close to a ground state and to only be significant in non-adiabatic situations. In the next paragraph we consider the case of two-electron spin-singlet states, for which Eq.~\eqref{eq:Kohn--Sham_geometric_corrected} can be solved explicitly for some choice of  $V_{\rm ext}+V^{\rm app}_{\rm Hxc}$.

\smallskip
\paragraph*{Two-electron spin-singlet.} 
A two-electron system in a singlet state has  
$\Phi(t,\br_1,\br_2)=\varphi(t,\br_1)\varphi(t,\br_2)( \ket{\uparrow\downarrow} - \ket{\downarrow\uparrow})
/\sqrt 2$.  All the previous variants of the modified KS equation can be written in the general form $i\partial_t\varphi=(-\nabla^2/2+V+iW)\varphi$ for some $V$ and $W$ chosen to recover the exact density $\rho^{\rm S}$, here assumed to be known. Writing $\varphi=\sqrt{\rho^{\mathrm{S}}/2} \, e^{i\theta}$, we see that $V, W, \theta$ solve 
\begin{subequations}\label{eq:hydro_singlet}
\begin{align}
W&=\frac{\partial_t\rho^{\rm S}+\nabla\cdot (\rho^{\rm S}\nabla\theta)}{2\rho^{\rm S}}, \label{eq:hydro_singlet_a}\\
V&=-\partial_t\theta-\frac{|\nabla\theta|^2}2+\frac{\Delta\sqrt{\rho^{\rm S}}}{2\sqrt{\rho^{\rm S}}}.
\label{eq:hydro_singlet_b}
\end{align}
\end{subequations}
In standard KS theory where $W=0$, $\theta$ is obtained by solving \eqref{eq:hydro_singlet_a}, and $V$ is given by \eqref{eq:hydro_singlet_b} with that $\theta$. For the geometric KS equation~\eqref{eq:Kohn--Sham_geometric_commutator} where $V = V_{\rm ext}$ the situation is reversed: One should first solve~\eqref{eq:hydro_singlet_b} to find $\theta$
and then $W$ is given by~\eqref{eq:hydro_singlet_a}. 
There are many solutions to \eqref{eq:hydro_singlet}. An important observation is that they can be parametrized by $\theta$, leading to explicit formulas for $V$ and $W$ from \eqref{eq:hydro_singlet}. The simplest case $\theta\equiv0$ yields the solution
\begin{equation}\label{eq:singlet-KS+G-potential}
    V^{\text{eaKS}} 
    = \frac{\Delta\sqrt{\rho^{\mathrm S}}}{2\sqrt{\rho^{\mathrm S}}},
    \qquad 
    W^{\text{na}} 
    =\frac{\partial_t\rho^{\mathrm S}}{2\rho^{\mathrm S}}.
\end{equation} 
The potential $V^{\rm eaKS}$ is the \emph{exact adiabatic KS potential}, defined as the total potential for which the time-dependent KS system has ground-state density $\rho^{\rm S}(t)$ for all $t$, and $W^{\text{na}}$ is the corresponding non-adiabatic geometric correction. We here apply the adiabatic approximation to the total \emph{exact KS potential} $V^{\rm eKS}:=V_0[\Phi_0,\rho^{\rm S}]$. It is not the usual adiabatic approximation, normally applied only to the Hxc potential in~\eqref{eq:V_HXC} \cite[Sec.~VIII.D]{MAQUI_finite-26_ppt}. It is remarkable that $W^{\text{na}}$ describing all the non-adiabatic effects is such a simple and explicit functional of $\rho^{\rm S}$.

\smallskip
\paragraph*{1D soft-Coulomb systems.}
We next consider a two-electron singlet in 1D, with a soft-Coulomb interaction $V_\text{ee}(x)=1/\sqrt{x^2+1}$. This framework was extensively studied~\cite{FukHelTokRub-11,HelFukCasVerMarTokRub-11,HelFukTokAppGroRub-11,EllFukRubMai-12,FukEllRubMai-13,DarBarMai-24}, because standard adiabatic TDDFT approximations fail to reproduce, even qualitatively, the exact $\rho^{\rm S}(t)$ for this 1D toy model in much the same way as they do for real 3D Coulombic systems. We consider two different time-dependent external potentials, both of the form $V_{\rm ext}^{(m)}(x)+xE\sin(\omega t)$.

First, as in~\cite{FukHelTokRub-11,DarBarMai-24}, we study Rabi oscillations in a Helium-like atom perturbed by an electric dipole interaction, corresponding to
$$V_{\rm ext}^{(1)}(x)=-\frac{2}{\sqrt{x^2+1}}$$
where $E=0.00667$ and $\omega= 0.5336$ is resonant with the first singlet excitation. The Rabi period is $T_{\rm R}=2\pi/(|d_{\rm eg}|E)$, where $d_{\rm eg}=\langle\Psi_{\rm e}|\hat{x}_1+\hat{x}_2|\Psi_{\rm g}\rangle$ and $\Psi_{\rm g}$ and $\Psi_{\rm e}$ are the ground and first excited states of $V_{\rm ext}^{(1)}(x)$.

To find a good approximation to $\rho^{\rm S}$, we solve a two-level time-dependent Schr\"odinger equation in the basis $(\Psi_{\rm g},\Psi_{\rm e})$. Those are obtained by diagonalizing a finite-difference approximation of the two-electron Hamiltonian at time $t=0$. The non-adiabatic contribution $V^{\text{na}}$ to the time-dependent exact KS potential is computed from \eqref{eq:hydro_singlet} with $W=0$. In 1D, we have~\cite{He2002,Mai-JCP-16}
\begin{equation*}
 V^{\text{na}}:= V^{\rm eKS}-V^{\rm eaKS}
 =- \frac{1}{2}\left(\frac{j_\varphi}{\rho^{\rm S}}\right)^2-\int^x\partial_t\left(\frac{j_\varphi}{\rho^{\rm S}}\right), 
\end{equation*}
where $j_\varphi=\rho^{\rm S}\nabla\theta$ is the current. This potential is compared to the geometric non-adiabatic term $W^{\rm na}$ in~\eqref{eq:singlet-KS+G-potential}. Our code is written in Julia and is freely available at~\cite{Duez2025GeometricTDDFTCode}.

\begin{figure*}
\includegraphics[scale=0.41]{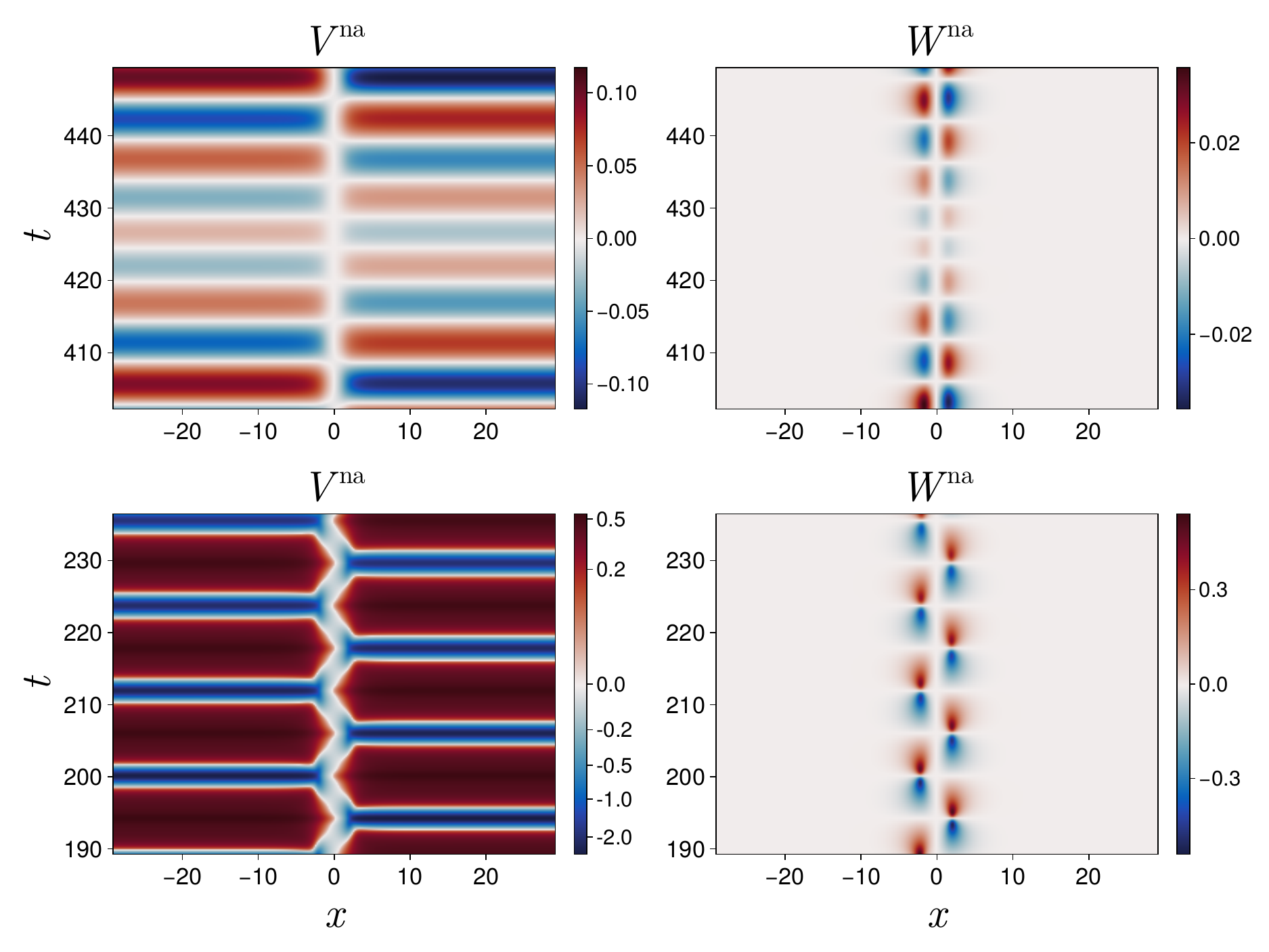}
\caption{Rabi oscillations in a 1D Helium-like atom with 
a uniform electric field of amplitude $E=0.00667$ and frequency $\omega=0.5336$. Heatmaps of the non-adiabatic contribution $V^\text{na}$ to the exact KS potential (left) and of the non-adiabatic geometric correction term $W^\text{na}$ (right), at times $|t-T_{\rm R}/4|\leq 2 T_{\rm opt}$ (bottom) and $|t-T_{\rm R}/2|\leq 2 T_{\rm opt}$ (top) with $T_{\rm opt}=2\pi/\omega$.
\label{fig:RabiOscillation}}
\includegraphics[scale=0.42]{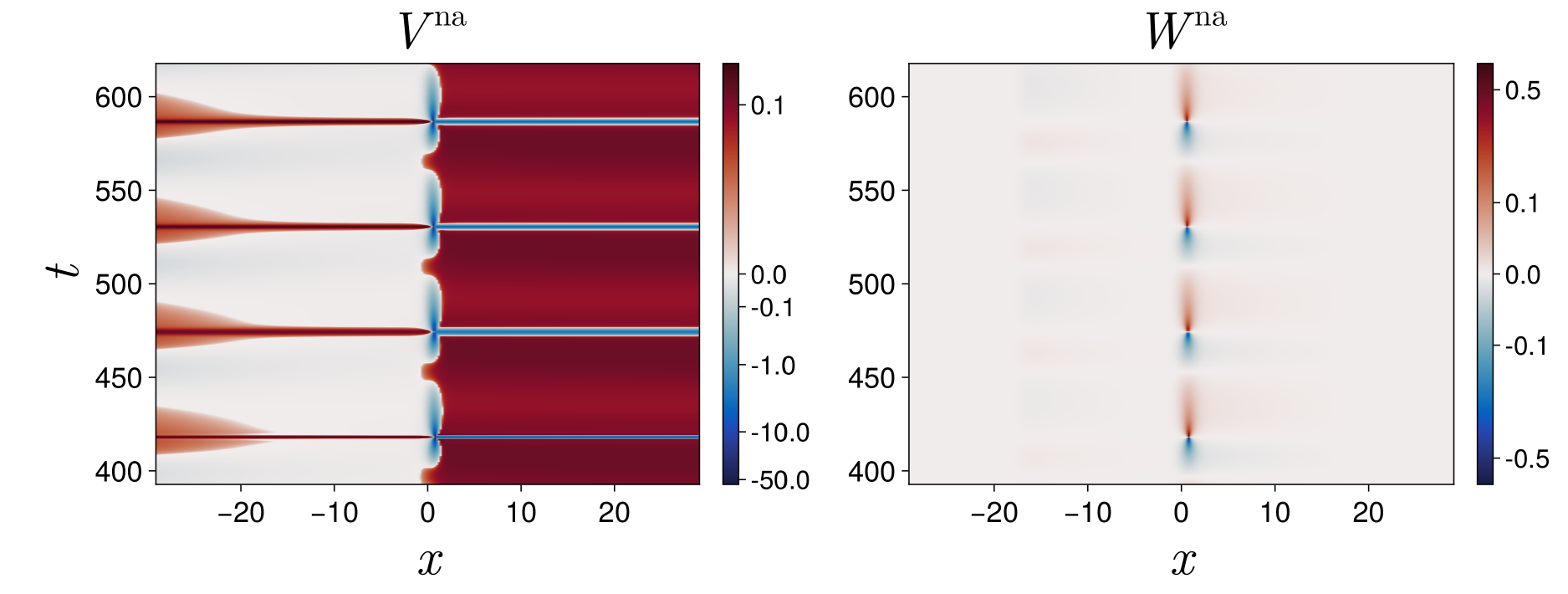}
\caption{Charge transfer between a 1D  Helium-like atom and a closed-shell-like atom with frozen electrons with 
a uniform electric field of amplitude $E=0.00667$ and frequency $\omega=0.112$. Heatmaps of the non-adiabatic contribution $V^\text{na}$ to the exact KS potential (left) and of the non-adiabatic geometric correction term $W^\text{na}$ (right), at times $|t-T_{\rm R}/8|\leq 2 T_{\rm opt}$ with $T_{\rm opt}=2\pi/\omega$. 
\label{fig:ChargeTransfert}}
\end{figure*}

In Fig.~\ref{fig:RabiOscillation}, we plot $V^{\rm na}$ and $W^{\rm na}$ over some intervals of times. Like in~\cite{EllFukRubMai-12}, we observe the presence of non-adiabatic fast-evolving steps in $V^{\text{na}}$ associated with a strong spatially nonlocal dependence on the density. In contrast, $W^{\rm na}$ is much more localized. Note also the scale difference between $V^{\text{na}}$ and $W^{\rm na}$.

Next, we consider a charge transfer from a Helium-type donor to a model of a closed-shell atom with frozen electrons, as in~\cite{FukEllRubMai-13}:
\begin{equation*}
    V_{\rm ext}^{(2)}(x)=-\frac{2}{\sqrt{\left(x+3.5\right)^2+1}}-\frac{1}{\cosh^2(x-3.5)}
\end{equation*}
with $E=0.00667$ and $\omega=0.112$. In Fig.~\ref{fig:ChargeTransfert}, we plot $V^{\text{na}}$ and $W^{\text{na}}$ over intervals of times. We find the same fast steps and high peaks in $V^{\rm na}$ as in~\cite{FukEllRubMai-13}. In contrast, $W^{\text{na}}$ is smaller and more localized, without high peaks.

\smallskip
\paragraph*{Conclusion.}
We have proposed a new TDDFT scheme based on geometric considerations. Our numerical simulations for the two-electron singlet suggest that approximating the non-adiabatic correction $W^{\rm na}$ could be easier than the corresponding non-adiabatic part of the standard KS potential, because it is a smoother function. Our new scheme could thus better describe out-of-equilibrium processes. In the companion article~\cite{MAQUI_finite-26_ppt} we study the mathematical foundations of our geometric approach for finite-dimensional systems. Future work will be devoted to constructing practical approximations of the geometric correction $W$ for real systems.

\smallskip
\paragraph*{Acknowledgments.}
This work has benefited from French State support managed by ANR under the France 2030 program through the MaQuI CNRS Risky and High-Impact Research program (RI)$^2$ (grant agreement ANR-24-RRII-0001). We thank the referees for useful comments.

%%%%%%%%%%%%%%%%%%%%%%%%%%%%%%%%%%%%%%%%%%%%%%%%%%%%%%%%
%%%%%%%%%%%%%%%%%%%%%%%%%%%%%%%%%%%%%%%%%%%%%%%%%%%%%%%%
%\bibliographystyle{aipauth4-2}
%\bibliography{biblio_maqui}

\appendix
\section{Geometric Runge--Gross theorem}\label{app:Runge--Gross}

The Runge--Gross theorem~\cite{RunGro-84} is a fundamental result in TDDFT. It asserts that two time-dependent external potentials giving rise to the same density must necessarily be equal, up to a time-dependent additive constant $C(t)$. Thus far, the mathematical proof of this result was only possible under the assumption that the time-dependent potentials are spatially smooth and analytic in time and that the initial wavefunction is smooth~\cite{FouLamLewSor-16}. The Runge--Gross argument therefore only applies to very smooth potentials such as the soft-Coulomb potential. Here we provide a rigorous proof that the same arguments as in~\cite{FouLamLewSor-16} apply to our geometric theory. More precisely, we work in space dimension $d\in\{1,2,3\}$ and, to cover all situations of interest, we give ourselves four functions $V_1(t,\br)$, $V_2(t,\br)$, $W_1(t,\br)$ and $W_2(t,\br)$. We will suppose that either $V_1=V_2$ or $W_1=W_2$ and want to deduce the missing equality. We assume that
\begin{itemize}
\itemsep0em
    \item[(i)] the interaction $\bV_{\rm ee}$ is infinitely differentiable in space and all its derivatives are bounded;
    \item[(ii)] the functions $V_1(t,\br)$, $V_2(t,\br)$, $W_1(t,\br)$ and $W_2(t,\br)$ are infinitely differentiable in space-time, with bounded space derivatives for all $t$;
    \item[(iii)] the initial state $\Psi_0$ is infinitely differentiable in space and all its derivatives are square-integrable.
\end{itemize}
We also need a further assumption that will be different depending on the situation. We will require that the initial state $\Psi_0$ satisfies one of the two conditions:
\begin{itemize}
\itemsep0em
    \item[(iv)] $\rho_{\Psi_0}(\br)>0$ everywhere; 
    \item[(iv')] the \emph{unique $V$-representability property}
\begin{equation}
\sum_{j=1}^NV(\br_j)\Psi_0(\br_1,...,\br_N)=0\quad \Longrightarrow\quad V(\br)=0
\label{eq:unique-V-rep}
\end{equation}
for all $V$ such that $\int_{\R^d}V(\br)^2\rho_{\Psi_0}(\br)\,\rd \br<\ii$. 
\end{itemize}
The equalities in~\eqref{eq:unique-V-rep} are understood Lebesgue-almost everywhere. The assumptions (i)--(iv) are from~\cite{FouLamLewSor-16}. The unique $V$-representability property~\eqref{eq:unique-V-rep} in (iv') is slightly stronger than (iv) and is further studied in~\cite{PenLeu-21} and in~\cite[Sec.~VII.B]{MAQUI_finite-26_ppt}. This is the main property used in the proof of the Hohenberg--Kohn theorem for ground-state DFT, where it follows from the unique continuation principle~\cite{Garrigue-20,LewLieSei-23_DFT}. In particular, both (iv) and (iv') are valid if $\Psi_0$ is the ground state of a Hamiltonian of the above form. For the theorem we do not need that $\Psi_0$ is a ground state but we do assume these conditions. As explained in~\cite{MAQUI_finite-26_ppt}, (iv') is the main property necessary for the geometric interpretation of TDDFT. 

\begin{theorem}[Geometric Runge--Gross]\label{thm:RGW}
Let $\Psi_1(t)$ and $\Psi_2(t)$ be the solutions to the Schrödinger equations
$$i\partial_t\Psi_m(t)=\bigg(\bT+\hat{V}_{\rm ee}+\sum_{j=1}^NV_m(t,\br_j)+iW_m(t,\br_j)\bigg)\Psi_m(t)$$
with $m=1$ or $m=2$, starting at the same initial condition $\Psi_0$. We assume {\rm (i)--(iii)} and that $\rho_{\Psi_1(t)}=\rho_{\Psi_2(t)}$ for $t\in [0,T)$.
\begin{itemize}
    \item[\rm (a)] Assume that $W_1=W_2$, that {\rm (iv)} holds and that the function $t\mapsto V_2(t,\br)-V_1(t,\br)$ is real-analytic for any $\br\in\R^d$. Then $V_1(t,\br)=V_2(t,\br)+C(t)$ for all $\br\in\R^d$ and all $t\in[0,T)$.
    \item[\rm (b)] Assume that $V_1=V_2$, that {\rm (iv')} holds and that the function $t\mapsto W_2(t,\br)-W_1(t,\br)$ is real-analytic for any $\br\in\R^d$. Then $W_1(t,\br)=W_2(t,\br)$ for all $\br\in\R^d$ and all $t\in[0,T)$.
\end{itemize}
\end{theorem}

The statement implicitly assumes that $\Psi_m(t)$ stays normalized, that is, $\int_{\R^d}W_m(t,\br)\rho_{\Psi_m(t)}(\br)\,\rd\br=0$. If we use the commutator form as in~\eqref{eq:correction_geometric} then this implicit assumption is not necessary and uniqueness for $W$ holds only up to a time-dependent constant $C(t)$ like for $V$.

The proof of Theorem~\ref{thm:RGW} follows closely~\cite{FouLamLewSor-16} and we only explain it in case (b). Under the conditions (i)--(iii) and (iv'), we show that
\begin{equation}
\frac{\partial^\ell}{\partial t^\ell}W_1(0,\br)=\frac{\partial^\ell}{\partial t^\ell}W_2(0,\br)
\label{eq:RG_for_W_ell}
\end{equation}
for all $\ell$ and all $\br\in\R^d$. The conclusion then follows from the assumed analyticity in time. The smoothness of the potentials and $\Psi_0$ implies that we can freely differentiate Schrödinger's equation in time~\cite{FouLamLewSor-16}. We denote by $\hat{W}_m(t):=\sum_{j=1}^N W_m(t,\br_j)$ the many-body operators associated with $W_1$ and $W_2$. We also introduce the difference $W:=W_2-W_1$ which we want to prove vanishes, and the corresponding $\cW:=\cW_2-\cW_1$. We call $\rho(t,\br)$ the common density of $\Psi_1(t)$ and $\Psi_2(t)$. Since in (b) we assume $V_1=V_2$, we call $H(t)=\bT+\hat{V}_{\rm ee}+\cV_1(t)$ the common part of the Hamiltonian.

We then give ourselves a smooth time-independent function $U(\br)$, that we will specify later. Using again the notation $\cU:=\sum_{j=1}^N U(\br_j)$, we have
$$\int_{\R^d}U(\br)\rho(t,\br)\,\rd\br=\pscal{\Psi_m(t),\hat{U}\Psi_m(t)}.$$
Differentiating this relation once in time gives
\begin{multline}
\int_{\R^d}U(\br)\partial_t\rho(t,\br)\,\rd\br\label{eq:rho_diff_1}\\
=\Big\langle\Psi_m(t),\big(i[\bT,\cU]+\cU\cW_m(t)+\cW_m(t)\cU\big)\Psi_m(t)\Big\rangle.
\end{multline}
We evaluate~\eqref{eq:rho_diff_1} at time $t=0$ and recall that $\Psi_1(0)=\Psi_2(0)=\Psi_0$. Subtracting the two equations (for $m=1$ and $m=2$) leads to
$$\pscal{\Psi_0,(\cU\cW(0)+\cW(0)\cU)\Psi_0}=0$$
where we recall that $W(0,\br)=W_2(0,\br)-W_1(0,\br)$. It is natural to take $U(\br)=W(0,\br)$ so that the previous relation reduces to $\|\cW(0)\Psi_0\|^2=0$, that is, $\cW(0)\Psi_0=0$. From the unique $V$-representability property~\eqref{eq:unique-V-rep}, this implies $W(0,\br)=W_2(0,\br)-W_1(0,\br)=0$ almost everywhere, hence everywhere by continuity. This is the claimed result~\eqref{eq:RG_for_W_ell} for $\ell=0$.

Next we go on and differentiate once more~\eqref{eq:rho_diff_1}, still with a general function $U$. To simplify the expression, we denote by $\bL_m^{(1)}(t):=i[\bT,\cU]+\cU\cW_m(t)+\cW_m(t)\cU$ the Hermitian operator appearing in~\eqref{eq:rho_diff_1}. The second-order derivative equals
\begin{multline}
\int_{\R^d}U(\br)\partial^2_t\rho(t,\br)\,\rd\br=\pscal{\Psi_m(t),\bK^{(1)}_m(t)\Psi_m(t)}\\
+\pscal{\Psi_m(t),\big(\cU\partial_t\cW_m(t)+\partial_t\cW_m(t)\cU\big)\Psi_m(t)}
\label{eq:rho_diff_2}
\end{multline}
with the operator
$$\bK^{(1)}_m(t):=i[\bH(t),\bL^{(1)}_m(t)]+\bL_m^{(1)}(t)\cW_m(t)+\cW_m(t)\bL^{(1)}_m(t).$$
We again evaluate at time $t=0$ and use that $\bL^{(1)}_1(0)=\bL^{(1)}_2(0)$ and $\bK^{(1)}_1(0)=\bK^{(1)}_2(0)$ since $\cW_1(0)=\cW_2(0)$, as we have proved in the first step. We find
$$\pscal{\Psi_0,\big(\cU\partial_t\cW(0)+\partial_t\cW(0)\cU\big)\Psi_0}=0.$$
Taking $U(\br)=\partial_tW(0,\br)=\partial_tW_2(0,\br)-\partial_tW_1(0,\br)$ and using again~\eqref{eq:unique-V-rep} for that $U$, we arrive at the conclusion that $\partial_tW_2(0,\br)=\partial_tW_1(0,\br)$, which is~\eqref{eq:RG_for_W_ell} for $\ell=1$.

The proof goes on like this by induction. We have
\begin{multline}
\int_{\R^d}U(\br)\partial^\ell_t\rho(t,\br)\,\rd\br=\pscal{\Psi_m(t),\bK^{(\ell-1)}_m(t)\Psi_m(t)}\\
+\pscal{\Psi_m(t),\big(\cU\partial^\ell_t\cW_m(t)+\partial^\ell_t\cW_m(t)\cU\big)\Psi_m(t)}
\label{eq:rho_diff_ell}
\end{multline}
for some operator $\bK^{(\ell-1)}_m$ defined by induction as
\begin{multline*}
\bK^{(\ell-1)}_m(t):=\partial_t\bK^{(\ell-2)}_m(t)+i\big[\bH(t),\bL^{(\ell-1)}_m(t)\big]\\
+\cW_m(t)\bL^{(\ell-1)}_m(t)+\bL^{(\ell-1)}_m(t)\cW_m(t),
\end{multline*}
where
\begin{equation*}
\bL^{(\ell-1)}_m(t):=\bK^{(\ell-1)}_m(t)+\cU\partial^{\ell-1}_t\cW_m(t)+\partial^{\ell-1}_t\cW_m(t)\cU.
\end{equation*}
The important remark is that $\bK_m^{(\ell-1)}(0)$ and $\bL_m^{(\ell-1)}(0)$ only depend on $\bW_m(0)$ and the derivatives $\partial^{k}_t\cW_m(0)$ up to order $k=\ell-1$, which coincide by the induction hypothesis.

\end{document}